\documentclass[aps,prd,twocolumn,amsmath,amssymb,amsfonts,nofootinbib,superscriptaddress,altaffilletter]{revtex4-1}
\usepackage{graphicx}
\usepackage{hyperref}
\hypersetup{
  bookmarksopen=true
}
\usepackage{amssymb}
\usepackage{amsmath}
\usepackage{color}
\usepackage{supertabular}
\usepackage{dcolumn}

\def\EatH{Einstein@Home }

\def\ec{\textrm{~,}}

\def\fdot{f^{(1)}}
\def\GFA{\textrm{GFA}}

\def\sci#1#2{#1\times10^{#2}}

\def\RAJ{\textrm{RA}_{\textrm J2000}}
\def\DECJ{\textrm{DEC}_{\textrm J2000}}

\begin{document}

\title{ 
A search of the Orion spur for continuous gravitational waves using a "loosely coherent" algorithm on data from LIGO interferometers.                                                                                         
}




%
%
%

\begin{abstract}
We report results of a wideband search for periodic gravitational waves from isolated neutron stars within the Orion spur towards both the inner and outer regions of our Galaxy. As gravitational waves interact very weakly with matter, the search is unimpeded by dust and concentrations of stars. One search disk (A) is $6.87^\circ$ in diameter and centered on 
$20^\textrm{h}10^\textrm{m}54.71^\textrm{s}+33^\circ33'25.29''$, and the 
other (B) is $7.45^\circ$ in diameter and centered on 
$8^\textrm{h}35^\textrm{m}20.61^\textrm{s}-46^\circ49'25.151''$.
We explored the frequency range of 50-1500~Hz and frequency derivative from $0$ to $\sci{-5}{-9}$~Hz/s.
A multi-stage, {\it loosely coherent} search program allowed probing more deeply than
before in these two regions, while increasing coherence length with every stage.


Rigorous followup parameters have winnowed initial coincidence set to only 70 candidates, to be examined manually.
None of those 70 candidates proved to be consistent with an isolated gravitational wave emitter,
and 95\%\ confidence level upper limits were placed on continuous-wave strain amplitudes.
Near $169$~Hz we achieve our lowest 95\%\ CL upper limit on worst-case linearly polarized strain amplitude $h_0$ of $\sci{6.3}{-25}$, while at the high end of our frequency range we achieve 
a worst-case upper limit of 
$\sci{3.4}{-24}$
for all polarizations and sky locations. 


\end{abstract}

%
%

\author{%
J.~Aasi,$^{1}$
B.~P.~Abbott,$^{1}$
R.~Abbott,$^{1}$
T.~D.~Abbott,$^{2}$
M.~R.~Abernathy,$^{1}$
F.~Acernese,$^{3,4}$
K.~Ackley,$^{5}$
C.~Adams,$^{6}$
T.~Adams,$^{7,8}$
P.~Addesso,$^{9}$
% ...
}\noaffiliation

\affiliation {LIGO---California Institute of Technology, Pasadena, CA 91125, USA }
\affiliation {Louisiana State University, Baton Rouge, LA 70803, USA }
\affiliation {Universit\`a di Salerno, Fisciano, I-84084 Salerno, Italy }
\affiliation {INFN, Sezione di Napoli, Complesso Universitario di Monte S.Angelo, I-80126 Napoli, Italy }
\affiliation {University of Florida, Gainesville, FL 32611, USA }
\affiliation {LIGO Livingston Observatory, Livingston, LA 70754, USA }
\affiliation {Cardiff University, Cardiff CF24 3AA, United Kingdom }
\affiliation {Laboratoire d'Annecy-le-Vieux de Physique des Particules (LAPP), Universit\'e Savoie Mont Blanc, CNRS/IN2P3, F-74941 Annecy-le-Vieux, France }
\affiliation {University of Sannio at Benevento, I-82100 Benevento, Italy and INFN, Sezione di Napoli, I-80100 Napoli, Italy }
\affiliation {Albert-Einstein-Institut, Max-Planck-Institut f\"ur Gravi\-ta\-tions\-physik, D-30167 Hannover, Germany }
\affiliation {Nikhef, Science Park, 1098 XG Amsterdam, The Netherlands }
\affiliation {LIGO---Massachusetts Institute of Technology, Cambridge, MA 02139, USA }
\affiliation {Instituto Nacional de Pesquisas Espaciais, 12227-010 S\~{a}o Jos\'{e} dos Campos, SP, Brazil }
\affiliation {Inter-University Centre for Astronomy and Astrophysics, Pune 411007, India }
\affiliation {International Centre for Theoretical Sciences, Tata Institute of Fundamental Research, Bangalore 560012, India }
\affiliation {University of Wisconsin-Milwaukee, Milwaukee, WI 53201, USA }
\affiliation {Leibniz Universit\"at Hannover, D-30167 Hannover, Germany }
\affiliation {Universit\`a di Siena, I-53100 Siena, Italy }
\affiliation {INFN, Sezione di Pisa, I-56127 Pisa, Italy }
\affiliation {Stanford University, Stanford, CA 94305, USA }
\affiliation {The University of Mississippi, University, MS 38677, USA }
\affiliation {California State University Fullerton, Fullerton, CA 92831, USA }
\affiliation {LAL, Universit\'e Paris-Sud, IN2P3/CNRS, F-91898 Orsay, France }
\affiliation {University of Southampton, Southampton SO17 1BJ, United Kingdom }
\affiliation {INFN, Sezione di Roma, I-00185 Roma, Italy }
\affiliation {Albert-Einstein-Institut, Max-Planck-Institut f\"ur Gravitations\-physik, D-14476 Golm, Germany }
\affiliation {Montana State University, Bozeman, MT 59717, USA }
\affiliation {Universit\`a di Perugia, I-06123 Perugia, Italy }
\affiliation {INFN, Sezione di Perugia, I-06123 Perugia, Italy }
\affiliation {European Gravitational Observatory (EGO), I-56021 Cascina, Pisa, Italy }
\affiliation {Syracuse University, Syracuse, NY 13244, USA }
\affiliation {SUPA, University of Glasgow, Glasgow G12 8QQ, United Kingdom }
\affiliation {LIGO Hanford Observatory, Richland, WA 99352, USA }
\affiliation {APC, AstroParticule et Cosmologie, Universit\'e Paris Diderot, CNRS/IN2P3, CEA/Irfu, Observatoire de Paris, Sorbonne Paris Cit\'e, F-75205 Paris Cedex 13, France }
\affiliation {Columbia University, New York, NY 10027, USA }
\affiliation {Universit\`a di Pisa, I-56127 Pisa, Italy }
\affiliation {CAMK-PAN, 00-716 Warsaw, Poland }
\affiliation {Astronomical Observatory Warsaw University, 00-478 Warsaw, Poland }
\affiliation {University of Birmingham, Birmingham B15 2TT, United Kingdom }
\affiliation {Universit\`a degli Studi di Genova, I-16146 Genova, Italy }
\affiliation {INFN, Sezione di Genova, I-16146 Genova, Italy }
\affiliation {RRCAT, Indore MP 452013, India }
\affiliation {Faculty of Physics, Lomonosov Moscow State University, Moscow 119991, Russia }
\affiliation {SUPA, University of the West of Scotland, Paisley PA1 2BE, United Kingdom }
\affiliation {University of Western Australia, Crawley, Western Australia 6009, Australia }
\affiliation {Department of Astrophysics/IMAPP, Radboud University Nijmegen, P.O. Box 9010, 6500 GL Nijmegen, The Netherlands }
\affiliation {ARTEMIS, Universit\'e Nice-Sophia-Antipolis, CNRS and Observatoire de la C\^ote d'Azur, F-06304 Nice, France }
\affiliation {MTA E\"otv\"os University, ``Lendulet'' Astrophysics Research Group, Budapest 1117, Hungary }
\affiliation {Institut de Physique de Rennes, CNRS, Universit\'e de Rennes 1, F-35042 Rennes, France }
\affiliation {Washington State University, Pullman, WA 99164, USA }
\affiliation {Universit\`a degli Studi di Urbino 'Carlo Bo', I-61029 Urbino, Italy }
\affiliation {INFN, Sezione di Firenze, I-50019 Sesto Fiorentino, Firenze, Italy }
\affiliation {Embry-Riddle Aeronautical University, Prescott, AZ 86301, USA }
\affiliation {University of Oregon, Eugene, OR 97403, USA }
\affiliation {Laboratoire Kastler Brossel, UPMC-Sorbonne Universit\'es, CNRS, ENS-PSL Research University, Coll\`ege de France, F-75005 Paris, France }
\affiliation {VU University Amsterdam, 1081 HV Amsterdam, The Netherlands }
\affiliation {University of Maryland, College Park, MD 20742, USA }
\affiliation {Center for Relativistic Astrophysics and School of Physics, Georgia Institute of Technology, Atlanta, GA 30332, USA }
\affiliation {Laboratoire des Mat\'eriaux Avanc\'es (LMA), IN2P3/CNRS, Universit\'e de Lyon, F-69622 Villeurbanne, Lyon, France }
\affiliation {Universitat de les Illes Balears---IEEC, E-07122 Palma de Mallorca, Spain }
\affiliation {Universit\`a di Napoli 'Federico II', Complesso Universitario di Monte S.Angelo, I-80126 Napoli, Italy }
\affiliation {NASA/Goddard Space Flight Center, Greenbelt, MD 20771, USA }
\affiliation {Canadian Institute for Theoretical Astrophysics, University of Toronto, Toronto, Ontario M5S 3H8, Canada }
\affiliation {Tsinghua University, Beijing 100084, China }
\affiliation {University of Michigan, Ann Arbor, MI 48109, USA }
\affiliation {Universit\`a di Roma Tor Vergata, I-00133 Roma, Italy }
\affiliation {INFN, Sezione di Roma Tor Vergata, I-00133 Roma, Italy }
\affiliation {National Tsing Hua University, Hsinchu Taiwan 300 }
\affiliation {Charles Sturt University, Wagga Wagga, New South Wales 2678, Australia }
\affiliation {Caltech---CaRT, Pasadena, CA 91125, USA }
\affiliation {Pusan National University, Busan 609-735, Korea }
\affiliation {Australian National University, Canberra, Australian Capital Territory 0200, Australia }
\affiliation {Carleton College, Northfield, MN 55057, USA }
\affiliation {INFN, Gran Sasso Science Institute, I-67100 L'Aquila, Italy }
\affiliation {Universit\`a di Roma 'La Sapienza', I-00185 Roma, Italy }
\affiliation {University of Brussels, Brussels 1050, Belgium }
\affiliation {INFN, Sezione di Padova, I-35131 Padova, Italy }
\affiliation {Texas Tech University, Lubbock, TX 79409, USA }
\affiliation {University of Minnesota, Minneapolis, MN 55455, USA }
\affiliation {The University of Melbourne, Parkville, Victoria 3010, Australia }
\affiliation {The University of Texas at Brownsville, Brownsville, TX 78520, USA }
\affiliation {The University of Sheffield, Sheffield S10 2TN, United Kingdom }
\affiliation {Wigner RCP, RMKI, H-1121 Budapest, Konkoly Thege Mikl\'os \'ut 29-33, Hungary }
\affiliation {Montclair State University, Montclair, NJ 07043, USA }
\affiliation {Argentinian Gravitational Wave Group, Cordoba Cordoba 5000, Argentina }
\affiliation {Universit\`a di Trento, Dipartimento di Fisica, I-38123 Povo, Trento, Italy }
\affiliation {INFN, Trento Institute for Fundamental Physics and Applications, I-38123 Povo, Trento, Italy }
\affiliation {The Pennsylvania State University, University Park, PA 16802, USA }
\affiliation {University of Chicago, Chicago, IL 60637, USA }
\affiliation {University of Cambridge, Cambridge CB2 1TN, United Kingdom }
\affiliation {University of Szeged, D\'om t\'er 9, Szeged 6720, Hungary }
\affiliation {Tata Institute for Fundamental Research, Mumbai 400005, India }
\affiliation {Institute for Plasma Research, Bhat, Gandhinagar 382428, India }
\affiliation {American University, Washington, D.C. 20016, USA }
\affiliation {University of Massachusetts-Amherst, Amherst, MA 01003, USA }
\affiliation {University of Adelaide, Adelaide, South Australia 5005, Australia }
\affiliation {West Virginia University, Morgantown, WV 26506, USA }
\affiliation {Korea Institute of Science and Technology Information, Daejeon 305-806, Korea }
\affiliation {University of Bia{\l }ystok, 15-424 Bia{\l }ystok, Poland }
\affiliation {SUPA, University of Strathclyde, Glasgow G1 1XQ, United Kingdom }
\affiliation {IISER-TVM, CET Campus, Trivandrum Kerala 695016, India }
\affiliation {Northwestern University, Evanston, IL 60208, USA }
\affiliation {Institute of Applied Physics, Nizhny Novgorod, 603950, Russia }
\affiliation {Hanyang University, Seoul 133-791, Korea }
\affiliation {NCBJ, 05-400 \'Swierk-Otwock, Poland }
\affiliation {IM-PAN, 00-956 Warsaw, Poland }
\affiliation {Monash University, Victoria 3800, Australia }
\affiliation {Seoul National University, Seoul 151-742, Korea }
\affiliation {ESPCI, CNRS, F-75005 Paris, France }
\affiliation {Universit\`a di Camerino, Dipartimento di Fisica, I-62032 Camerino, Italy }
\affiliation {Southern University and A\&M College, Baton Rouge, LA 70813, USA }
\affiliation {College of William and Mary, Williamsburg, VA 23187, USA }
\affiliation {Instituto de F\'\i sica Te\'orica, University Estadual Paulista/ICTP South American Institute for Fundamental Research, S\~ao Paulo SP 01140-070, Brazil }
\affiliation {IISER-Kolkata, Mohanpur, West Bengal 741252, India }
\affiliation {Rutherford Appleton Laboratory, HSIC, Chilton, Didcot, Oxon OX11 0QX, United Kingdom }
\affiliation {Whitman College, 280 Boyer Ave, Walla Walla, WA 9936, USA }
\affiliation {National Institute for Mathematical Sciences, Daejeon 305-390, Korea }
\affiliation {Rochester Institute of Technology, Rochester, NY 14623, USA }
\affiliation {Hobart and William Smith Colleges, Geneva, NY 14456, USA }
\affiliation {Institute of Astronomy, 65-265 Zielona G\'ora, Poland }
\affiliation {Andrews University, Berrien Springs, MI 49104, USA }
\affiliation {Trinity University, San Antonio, TX 78212, USA }
\affiliation {Universit\`a di Padova, Dipartimento di Fisica e Astronomia, I-35131 Padova, Italy }
\affiliation {University of Washington, Seattle, WA 98195, USA }
\affiliation {Abilene Christian University, Abilene, TX 79699, USA }
\affiliation {Emory University, Atlanta, GA 30322, USA} 


%
%
\maketitle

\section{Introduction}
\label{sec:introduction}

In this paper we report the results of a deep search along the Orion spur for continuous, nearly monochromatic gravitational waves in data from LIGO's sixth science (S6) run. The search covered frequencies from 50~Hz through 1500~Hz and frequency derivatives from $0$~Hz/s through $-\sci{5}{-9}$~Hz/s.

Our solar system is located in the Orion spur --- a spoke-like concentration of stars connecting the Sagittarius and Perseus arms of our galaxy. 
Since known pulsars tend to be found in concentrations of stars such as galactic arms and globular clusters \cite{PARKES_paper, OBstars}, the Orion spur offers a potential target. This search explores a portion of the Orion spur towards the inner regions of our Galaxy as well as a nearly opposite direction covering the Vela nebula.

A number of searches have been carried out previously on LIGO data~\cite{S4IncoherentPaper, EarlyS5Paper, FullS5Semicoherent, S2TDPaper, S3S4TDPaper, S2FstatPaper, Crab, pulsars3, CasA}, including coherent searches for graviational waves from known radio and X-ray pulsars. An \EatH search running on the BOINC infrastructure \cite{BOINC} has performed blind all-sky searches on data from LIGO's S4 and S5 science runs~\cite{S4EH, S5EH, FullS5EH}.

The results in this paper were produced with the PowerFlux search code. It was first described in \cite{S4IncoherentPaper} together with two other semi-coherent search pipelines (Hough, Stackslide). The sensitivities of all three methods were compared, with PowerFlux showing better results in
frequency bands lacking severe spectral artifacts.
A subsequent article~\cite{FullS5Semicoherent} based on the data from the S5 run featured improved upper limits 
and a systematic follow-up detection search based on the {\em Loosely coherent} algorithm~\cite{loosely_coherent}.

In this paper we establish the most sensitive wide-band upper limits to date in the frequency band 50-1500 Hz. Near $169$~Hz our strain sensitivity to 
a neutron star with the most unfavorable sky location and orientation (``worst case'') yields a 95\% confidence level
upper limit in intrinsic strain amplitude of $\sci{6.3}{-25}$, while at the high end of our frequency range we achieve 
a worst-case upper limit of 
$\sci{3.4}{-24}$.

Starting from 94,000 outliers surviving the first stage of the pipeline, only
70 survived the fourth and final stage of the automated search program and
were then examined manually for instrumental contamination. Of the 70 outliers found, several do not have an easily identifiable instrumental cause.

Deeper follow-ups of the outliers
do not lead to increased statistical significance, as would be expected for
a GW-emitting isolated neutron star. Accurate
estimation of the probability for a statistical fluctuation to lead to the
loudest of these outliers, using simulation of the search on independent data sets, 
is computationally infeasible, but a rough (conservative) estimate (described in section \ref{sec:GFA}) is 
O(10\%). Given this modest improbability and given the inconsistency of
deep follow-up results with the isolated signal model, we conclude that 
statistical fluctuations are a likely explanation for these outliers.

As the deeper follow-up searches assumed a tight coherence length, this leaves open a narrow window for the outliers to be caused by neutron star with an additional frequency modulation such as would be observed if it were in long-period orbit. The enlargement of parameter-space needed to cover this possibility makes it impractical to test this hypothesis with S6 data. 


\section{LIGO interferometers and S6 science run}

The LIGO gravitational wave network consists of two observatories, one in Hanford, Washington and the other in Livingston, Louisiana, separated by a 3000-km baseline. During the S6 run each site housed one suspended interferometer with 4-km long arms. 

Although the sixth science run spanned more than one year period of data acquisition, the analysis in this paper used data only from GPS 951534120 (2010  Mar 02 03:01:45 UTC) through GPS 971619922 (2010 Oct 20 14:25:07 UTC), selected for good strain sensitivity and noise stationarity.  Since interferometers sporadically fall out of operation (``lose lock'') due to environmental or instrumental disturbances or for scheduled maintenance periods, the data set is not contiguous. For the time span used in the search the Hanford interferometer H1 had a duty factor of 53\%, while the Livingston interferometer L1 had a duty factor of 51\% . The strain sensitivity in search band was not uniform, exhibiting a $\sim 50$\% daily variation from anthropogenic activity as well as gradual improvement toward the end of the run \cite{LIGO_detector, detchar}.

A thorough description of instruments and data can be found in \cite{detchar2}.

\section{Search region}

\begin{table*}[htbp]
\begin{center}
\begin{tabular}{lrrrrrr}\hline
Search region & \multicolumn{1}{c}{RA} & \multicolumn{1}{c}{DEC} & \multicolumn{1}{c}{Radius} & \multicolumn{1}{c}{RA} & \multicolumn{1}{c}{DEC} & \multicolumn{1}{c}{Radius} \\
 & \multicolumn{1}{c}{rad} & \multicolumn{1}{c}{rad} & \multicolumn{1}{c}{rad} &\multicolumn{1}{c}{hours} & \multicolumn{1}{c}{deg} &\multicolumn{1}{c}{deg}  \\
\hline \hline \\

A & 5.283600 & 0.585700& 0.060 & $20^h10^m54.715^s$ & $33^\circ 33'29.297''$ & 3.438 \\
B & 2.248610 & -0.788476& 0.065 & $8^h35^m20.607^s$ & $-46^\circ 49'25.151''$ & 3.724 \\
\hline
\end{tabular}
\caption[Search region]{Area of sky covered by this search.}
\label{tab:search_region}
\end{center}
\end{table*}

All-sky searches for continuous gravitational waves in data produced by modern interferometers are computationally limited, with the established upper limits an order of magnitude away from what is theoretically possible given  impractically large computational resources. 
This limitation arises from the rapid increase in computational cost with coherence time of the search, 
because of both the necessarily finer gridding of the sky and the need to search over higher-order derivatives of the 
signal frequency.
Hence there is a tradeoff between searching the largest sky area with reduced sensitivity of all-sky search, and pushing for sensitivity in a smaller region.

The loosely coherent search program was initially developed for follow-up of outliers from an all-sky semicoherent search~\cite{FullS5Semicoherent}. For this search we have chosen to isolate two small regions and take advantage of the enhanced sensitivity of the loosely coherent search. Besides the gain from increasing coherence length we also benefit from search regions (listed in Table \ref{tab:search_region}) with strong Dopper-modulated frequency evolution and greater rejection of instrumental artifacts.

Known radio pulsars tend to cluster along the spiral arms, in globular clusters, and in other star-forming regions. 
To increase the chances of discovering a continuous wave gravitational source we selected regions where 
one can expect a clustering of neutron star sources in line-of-sight cones determined by the search area and sensitivity reach of the detector.

\begin{figure}[htbp]
\begin{center}
  \includegraphics[width=3.3in]{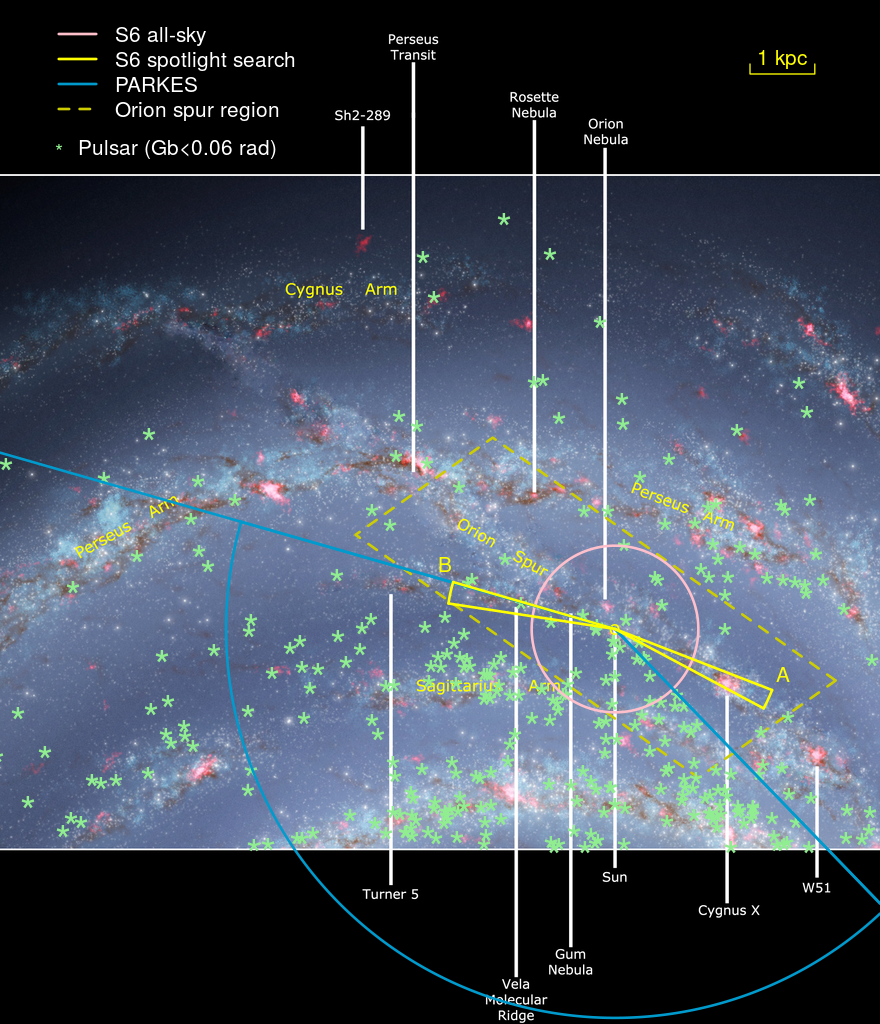}
  
 \caption{Distribution of known pulsars in the Milky Way galaxy. The Orion spur region (marked by dashed rectangle) connects Perseus and Sagittarius galactic arms and includes regions marked A and B. The ranges shown for gravitational wave searches correspond to 1500~Hz frequency and an ellipticity of $10^{-6}$. The arc shown for the PARKES survey \cite{PARKES_paper} shows search area, not the range. The green stars show locations of pulsars from the ATNF database (retrieved on January 29, 2015, \cite{ATNF}) with galactic latitude $\textrm{Gb}$ below $0.06$ radians. The background image is due to R.~Hurt \cite{milkiway} (color online)}
\label{fig:milkyway_pulsars_local}
\end{center}
\end{figure}

The positions of known pulsars from the ATNF catalog (\cite{ATNF, ATNFURL}, retrieved 2015 Jan 29) and the expected reach of semicoherent searches are illustrated in Figure \ref{fig:milkyway_pulsars_local} on the galactic background \cite{milkiway}. Only pulsars with galactic latitude less than $0.06$~rad are shown in the figure. We observe loose association with galactic arms, which is skewed by observational bias. In particular, the area searched by Parkes survey marked as a blue sector contains many more pulsars than elsewhere on the map. 

The expected reach of the all-sky search in S6 data, assuming a neutron star ellipticity of $10^{-6}$, is illustrated by the pink circle. A computationally feasible spotlight search can reach twice as far, but the globular clusters and galactic center remain out of its reach in the S6 data set.

A closer alternative is to look in the local neighbourhood of the Sun along the Orion spur --- a grouping of stars that connects the Perseus and Sagittarius arms of our galaxy. For this search we have chosen two regions (Table \ref{tab:search_region}), exploring two nearly opposite directions along the Orion spur. 

Region A was chosen to point near Cygnus X, with region B pointing toward the Vela nebula \ref{tab:search_region}. A recent study of
OB stars and their ramificatons for local supernova rates support these two directions as potentially promising, along with several 
other star-forming regions~\cite{OBstars}. The choice of sky area to search for region B is more ambiguous because of larger extent of Orion spur --- the figure \ref{fig:milkyway_pulsars_local} shows two grouping of stars towards the Vela Molecular Ridge and Perseus transit directions. We have chosen the direction towards Vela as it coincides with star forming region with several known neutron stars. In order to better cover Vela nebula the region B search radius is slightly larger than that of region A.


%
%
%

\section{Search algorithm}

The results presented in this paper were obtained with the loosely coherent search, implemented as part of the PowerFlux program. We have used the follow-up procedure developed for the all-sky S6 search, but where the first loosely coherent stage is applied directly to the entire A and B regions.
A detailed description of the loosely coherent code can be found in \cite{FullS5Semicoherent, loosely_coherent}. 

Mathematically, we transform the input data to the Solar System barycentric reference frame, correct for putative signal evolution given by frequency, spindown and polarization parameters, and then apply a low-pass filter which bandwidth determines the coherence length of the search. The total power in the computed time series is then compared to power obtained for nearby frequency bins in a 0.25~Hz interval.

A signal-to-noise ratio and an upper limit are derived for each frequency bin using a universal statistic method \cite{universal_statistics} that establishes 95\% CL upper limit for an arbitrary underlying noise distribution. If the noise is Gaussian distributed the upper limits are close to optimal values that would be produced with assumption of Gaussianity. For non-Gaussian noise the upper limits are conservatively correct.

\begin{figure*}[htbp]
\begin{center}
  \includegraphics[width=7.2in]{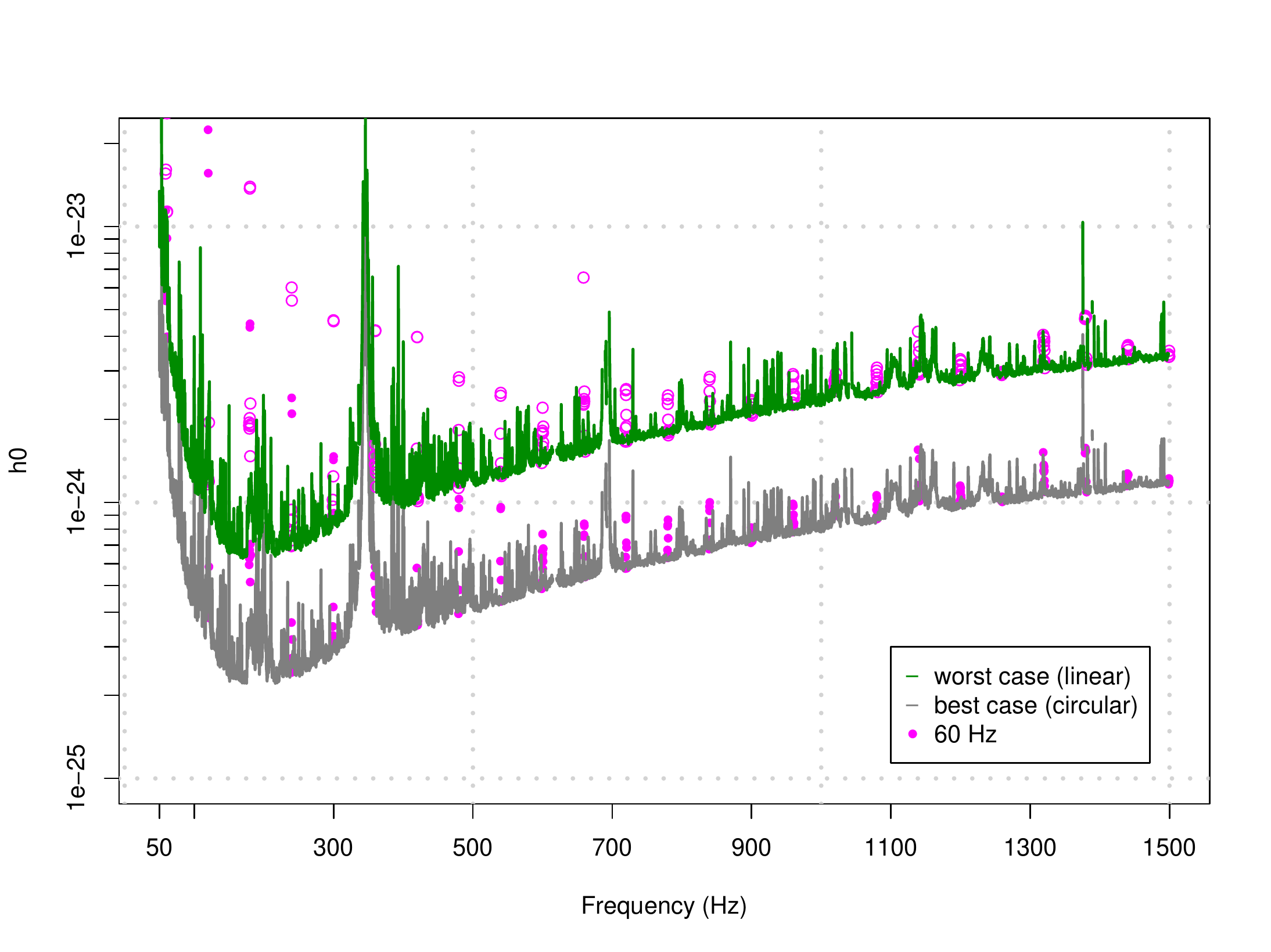}

 \caption[Upper limits]{S6 95\%\ CL upper limits on signal strain amplitude. The upper (green) curve shows worst case upper limits in analyzed 0.25~Hz bands (see Table \ref{tab:excluded_ul_bands} for list of excluded bands). The lower (grey) curve shows upper limits assuming circularly polarized source. The values of solid points and circles (marking power line harmonics for circularly and linear polarized sources) are not considered reliable. They are shown to indicate contaminated bands. (color online) }
\label{fig:full_s6_upper_limits}
\end{center}
\end{figure*}

\begin{figure}[htbp]
\includegraphics[width=3in]{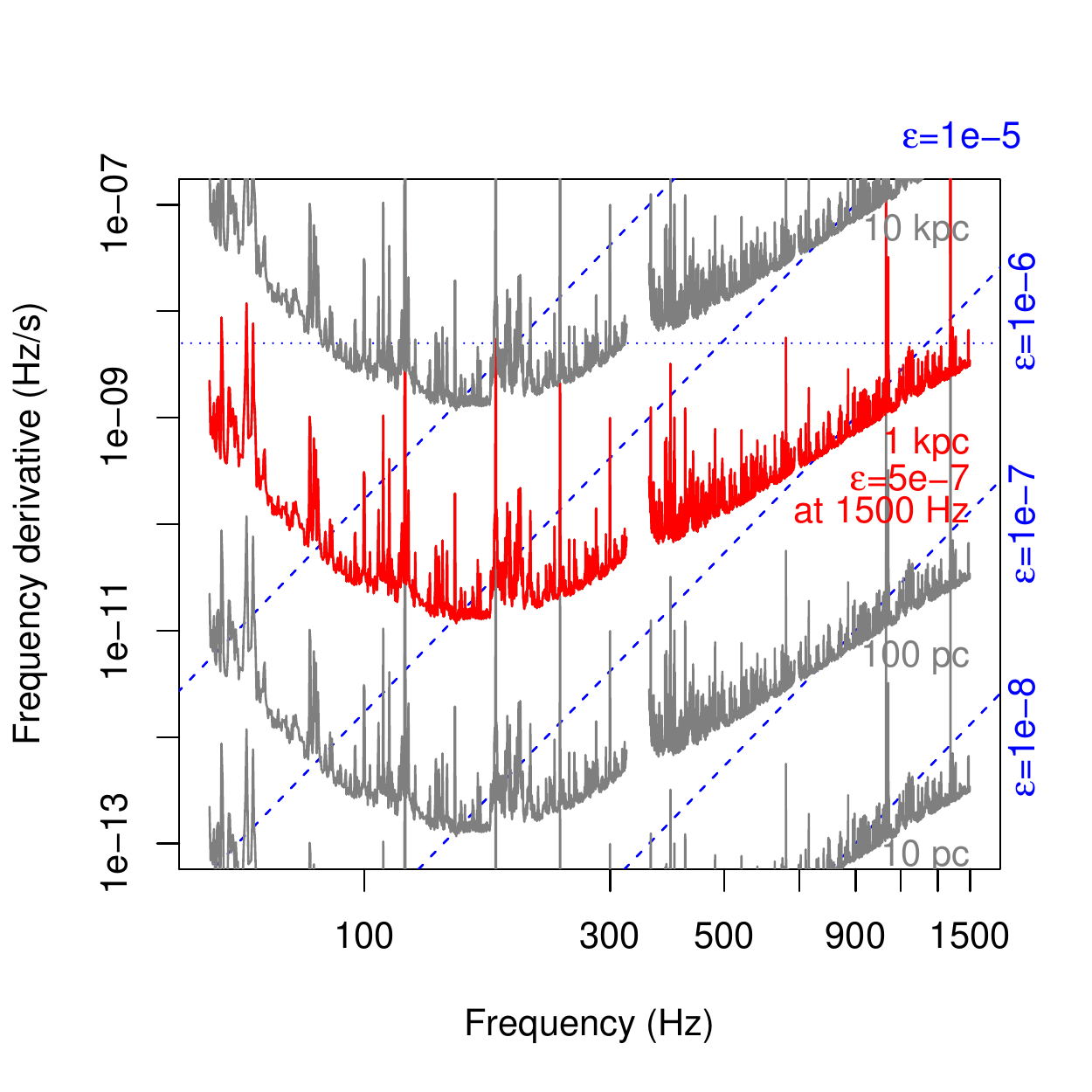}
\caption[Spindown range]{
\label{fig:spindown_range}
Range of the PowerFlux search for neutron stars
spinning down solely due to graviational waves.  This is a
superposition of two contour plots.  The grey and red solid lines are contours of the maximum distance at which a neutron
star in optimum orientation could be detected as a function of gravitational-wave frequency
$f$ and its derivative $\dot{f}$.  The dashed lines 
are contours of the corresponding ellipticity
$\epsilon(f,\dot{f})$. The fine dotted line marks the maximum spindown searched. Together these quantities tell us the
maximum range of the search in terms of various populations (see text
for details). In particular, at 1500 Hz we are sensitive to stars with ellipticity of $\sci{5}{-7}$ up to $1$~kpc away. (color online) 
}
\end{figure}

Maxima of the SNR and upper limits over marginalized search parameters are presented in the plots \ref{fig:full_s6_upper_limits}, \ref{fig:spindown_range} and \ref{fig:snr_skymap14_wide}.

The search results described in this paper assume a classical model of a spinning neutron star with a fixed, non-axisymmetric mass quadrupole that produces circularly polarized graviational waves along the rotation axis and linearly polarized radiation in the directions perpendicular to the rotation axis.
The assumed signal model is thus
\begin{equation}
\begin{array}{l}
h(t)=h_0\left(F_+(t, \alpha, \delta, \psi)\frac{1+\cos^2(\iota)}{2}\cos(\Phi(t))+\right.\\
\quad\quad\quad \left.\vphantom{\frac{1+\cos^2(\iota)}{2}}+F_\times(t, \alpha, \delta, \psi)\cos(\iota)\sin(\Phi(t))\right)\ec
\end{array}
\end{equation}

\noindent where $F_+$ and $F_\times$ characterize the detector responses to signals with ``$+$'' and ``$\times$'' 
quadrupolar polarizations, the sky location is described by right ascension $\alpha$ and declination $\delta$, $\iota$ describes the inclination of the source rotation axis to the line of sight, and the phase evolution of the signal is given by the formula
\begin{equation}
\label{eqn:phase_evolution}
\Phi(t)=2\pi(f_\textrm{source}(t-t_0)+\fdot(t-t_0)^2/2)+\phi\ec
\end{equation}
with $f_\textrm{source}$ being the source frequency and $\fdot$ denoting the first frequency derivative (for which we also use the abbreviation {\em spindown}). $\phi$ denotes the initial phase with respect to reference time $t_0$. $t$ is time in the solar system barycenter frame. When expressed as a function of local time of ground-based detectors it includes the sky-position-dependent Doppler shift.
We use $\psi$ to denote the polarization angle of the projected source rotation axis in the sky plane.

As a first step, individual SFTs (short Fourier transforms) with high noise levels or large spikes in the underlying data are removed from the analysis. For a typical well-behaved frequency band, we can exclude 8\% of the SFTs while losing only 4\% of the accumulated statistical weight.
For a band with large detector artifacts (such as instrumental lines arising from resonant
vibration of mirror suspension wires), however, we can end up removing most, if not all, SFTs. As such bands are not expected to have any sensitivity of physical interest they were excluded from the upper limit analysis (Table \ref{tab:excluded_ul_bands}).

\begin{table}[htbp]
\begin{center}
\begin{tabular}{lp{3.5cm}}\hline
Category & Description \\
\hline \hline
60 hz line & 59.75-60.25~Hz \\
Violin modes & 343.25-343.75~Hz, 347~Hz \\
Second harmonic of violin modes & 687.00-687.50~Hz\\
Third harmonic of violin modes & 1031.00-1031.25~Hz \\
\hline
\end{tabular}
\caption[Frequency regions excluded from upper limit analysis]{Frequency regions excluded from upper limit analysis. These are separated into power line artifacts and harmonics of ``violin modes'' (resonant vibrations of the wires which suspend the many mirrors of the interferometer).}
\label{tab:excluded_ul_bands}
\end{center}
\end{table}

The detection pipeline used in this search was developed for an S6 all-sky analysis
and is an extension of the pipeline described in~\cite{FullS5Semicoherent}. It consists of several stages employing loosely coherent \cite{loosely_coherent} search algorithm with progressively stricter coherence requirements. The parameters of the pipeline are described in Table~\ref{tab:followup_parameters}. 

Unlike in the all-sky analysis the first stage is used to establish upper limits. In effect, instead of investigating all-sky outliers we have simply pointed the follow-up pipeline along the direction of Orion spur. This allowed us to increase the sensitivity by a factor of $2$. The rest of the pipeline is unmodified.

The frequency refinement parameter is specified relative to the $1/1800$~Hz frequency bin width used in SFTs that serve as input to the analysis.
Thus at the last stage of follow-up our frequency resolution is $(1800\textrm{~s} \cdot 32)^{-1}=17$~$\mu$Hz. However, because of the degeneracy between frequency, sky position and spindown, the accuracy is not as good and the frequency can deviate by up to $50$~$\mu$Hz in $95\%$ of injections. 
This degeneracy is mostly due to Doppler shifts from Earth orbital motion and is thus common to both interferometers.

The phase coherence parameter $\delta$ is described in detail in \cite{loosely_coherent}. It represents the amount of allowed phase variation over a $1800$~s interval. We are thus sensitive both to the expected sources with ideal frequency evolution (equation \ref{eqn:phase_evolution}) and unexpected sources with a small amount of frequency modulation.

The sky refinement parameter is relative to the sky resolution sufficient for the plain semi-coherent PowerFlux mode and was necessary because  the improved frequency resolution made the search more sensitive to Doppler shift.

\begin{table*}[htbp]
\begin{center}
\begin{tabular}{llccccc}\hline
Stage & Instrument sum & {Phase coherence} & \multicolumn{1}{c}{Spindown step} & \multicolumn{1}{c}{Sky refinement} & \multicolumn{1}{c}{Frequency refinement} & \multicolumn{1}{c}{SNR increase} \\
 & & \multicolumn{1}{c}{rad} & \multicolumn{1}{c}{Hz/s} &  &  & \multicolumn{1}{c}{\%}\\
\hline \hline \\

  1 & incoherent & $\pi/2$ & $\sci{1.0}{-10}$ & $1/4$ & $1/8$ & NA\\
  2 & coherent & $\pi/2$ & $\sci{5.0}{-11}$ & $1/4$ & $1/8$ & 0  \\
  3 & coherent & $\pi/4$ & $\sci{2.5}{-11}$ & $1/8$ & $1/16$ & 12 \\
  4 & coherent & $\pi/8$ & $\sci{5.0}{-12}$ & $1/16$ & $1/32$ & 12 \\
 \hline
\end{tabular}
\caption[Analysis pipeline parameters]{Analysis pipeline parameters. All stages used the loosely coherent algorithm for demodulation. The sky and frequency refinement parameters are relative to values used in the semicoherent PowerFlux search.}
\label{tab:followup_parameters}
\end{center}
\end{table*}

Stages one and two used the same parameters, with the only difference being that data acquired at nearby times by different interferometers were combined without regard to phase in stage 1, but we took phase into account in stage 2. In the ideal situation, when both detectors are operational at the same time and at the same sensitivity, one would expect an increase in SNR by $\sqrt{2}$ by including phase information. In practice, the duty cycle did not overlap perfectly and, most importantly, it was quite common for one interferometer to be more sensitive than another. Thus, to keep an outlier, we only required that SNR did not decrease when transitioning to stage 2.

Subsequent stages used longer coherence times, with correspondingly finer sky and frequency resolutions.

The analysis data set was partitioned in time into 7 parts of equal duration numbered 0 through 6. As an intermediate product we have obtained upper limits and outliers of each contiguous sequence of parts. For example, a segment [1,5] would consist of the middle 5/7 of the entire data set.
This allowed us to identify outliers that exhibited enhanced SNR on a subset of data and thus were more likely to be induced by instrumental artifacts (Tables \ref{tab:outliersA} and \ref{tab:outliersB}).

\section{Gaussian false alarm event rate}
\label{sec:GFA}

The computation of the false alarm rate for the outliers passing all stages of the pipeline is complicated by the fact that most outliers are caused by instrumental artifacts for which we do not know the underlying probability distribution. In principle, one could repeat the analysis many times using non-physical frequency shifts (which would exclude picking up a real signal by accident) in order to obtain estimates of false alarm rate, but this approach incurs prohibitive computational cost.
Even assuming a perfect Gaussian background, it is difficult to model the pipeline in every detail to obtain an accurate estimate of the false alarm rate, given the gaps in interferometer operations and non-stationary noise.

Instead, we compute a figure of merit that overestimates the actual Gaussian false alarm event rate. We simplify the problem by assuming that the entire analysis was carried out with the resolution of the very last stage of follow-up and we are merely triggering on the SNR value of the last stage.
This is extremely conservative as we ignore the consistency requirements that allow the outlier to proceed from one stage of the pipeline to the next, actual false alarm rate could be lower.

The SNR of each outlier is computed relative to the loosely coherent power sum for 501 frequency bins spaced at $1/1800$~Hz intervals (including the outlier) but with all the other signal parameters held constant. The spacing assures that any sub-bin leakage does not affect the statistics of the power sum. 

As the power sums are weighted, the statistics should follow a weighted $\chi^2$ distribution, the exact shape of which is difficult to characterize analytically because the weights depend on sky position, gaps in acquired data, background noise in the SFTs and the polarization parameters of the outlier.

To simplify computation we  assume that we are dealing with a simple $\chi^2$ distribution with the number of degrees of freedom given by the timebase divided by the coherence length and multiplied by a conservative duty factor reflecting interferometer uptime and the worst-case weights from linearly-polarized signals.

Thus to find the number of degrees of freedom we will use the formula

\begin{equation}
\label{N_chi2}
N\approx \frac{\textrm{timebase} \cdot \delta  \cdot \textrm{duty factor}}{ 1800\textrm{~s} \cdot 2\pi }
\end{equation}
with the duty factor taken to be $0.125$ and $\delta$ giving the phase coherence parameter of the loosely coherent search. The duty factor was chosen to allow for only $50$\% interferometer uptime and only one quarter of the data receiving high weights from our weighting scheme, which weights the contribution of data inversely as the square of the estimated noise \cite{PowerFluxTechNote, PowerFlux2TechNote}.

The number of search templates that would be needed if the last stage of follow-up were used on the entire search region is conservatively (over)estimated as
\begin{equation}
K=\sci{5.8}{7} \frac{f_1^3-f_0^3}{1400.25^3-1400^3} 
\end{equation}
where $f_0$ and $f_1$ (in Hz) describe the frequency band of interest. For any particular $0.25$~Hz search band the number of templates scales quadratically in frequency due to linearly growing influence of Doppler shifts. Thus the integrated frequency dependence is cubic. The scaling factor $\sci{5.8}{7}$ was obtained by counting the number of templates for a particular PowerFlux instance that searched from $1400$~Hz to $1400.25$~Hz. For the entire analysis $f_0=50$~Hz and $f_1=1500$~Hz, which yields $K=\sci{1.3}{11}$ templates, without accounting for template overlap.  

Thus we define the outlier figure of merit describing Gaussian false alarm event rate as 
\begin{equation}
\label{GFA}
\GFA = K\cdot P_{\chi^2}\left(N+\textrm{SNR}\cdot \sqrt{2 N} ; N\right)
\end{equation}
where $N$ defines the number of degrees of freedom as given by equation \ref{N_chi2}, $P_{\chi^2}(x ; N)$ gives the probability for a $\chi^2$ distribution with $N$ degrees of freedom to exceed $x$, and $K$ describes the estimated number of templates.

We point out that the $\GFA$ is overly conservative when applied to frequency bands with Gaussian noise, but is only loosely applicable to bands with detector artifacts, which can affect both the estimate of the degrees of freedom of the underlying distribution and the assumption of uncorrelated underlying noise.


%
%

\section{Results}
\label{sec:results}

\begin{table}[htbp]
\begin{center}
\begin{tabular}{lrr}\hline
Stage & Region A & Region B  \\
 & &   \\
\hline \hline \\

1 & 43884 & 51027 \\
2 & 7921 & 9152  \\
3 & 510 & 566  \\
4 & 37 & 33  \\
\hline
\end{tabular}
\caption[Outlier counts]{Outlier counts found at each stage of follow up.}
\label{tab:outlier_counts}
\end{center}
\end{table}

PowerFlux produces 95\% confidence level upper limits for individual templates, where each template                                                                                               
represents a particular value of frequency, spindown, sky location and polarization. The results are maximized over several parameters, and a correction factor is applied to account for possible mismatches between a true signal and sampled parameters. Figure \ref{fig:full_s6_upper_limits} shows the resulting upper limits maximized over the analyzed spindown range, over the search regions and, for the upper curve, over all sampled polarizations. The lower curve shows the upper limit for circular polarized signals alone.

The numerical data for this plot can be obtained separately \cite{data}.

The regions near harmonics of the 60 Hz power mains frequency are shown as circles. 

Figure \ref{fig:spindown_range} provides an easy way to judge the astrophysical range of the search. We have computed the implied spindown solely due to gravitational emission at various distances, as well as corresponding ellipticity curves, assuming a circularly polarized signal. This follows formulas in paper \cite{S4IncoherentPaper}. For example, at the highest frequency sampled, assuming ellipticity of $\sci{5}{-7}$ (which is well under the maximum limit in \cite{crust_limit, crust_limit2}) we can see as far as $1000$ parsecs.

In each search band, including regions with detector artifacts, the follow-up pipeline  was applied to outliers satisfying the initial                                                                                                  
coincidence criteria. The outlier statistics are given in Table \ref{tab:outlier_counts}. The outliers that passed all stages of the automated pipeline are listed in Table \ref{tab:outliersA} for the A direction and Table \ref{tab:outliersB} for the B direction. Each
of these outliers was inspected manually and tested against further
criteria to determine whether it was convincingly due to a source in
the targeted astrophysical population.

\begin{table*}[htbp]
\begin{center}
\small
\begin{tabular}{D{.}{.}{2}D{.}{.}{2}D{.}{.}{2}rD{.}{.}{5}rD{.}{.}{4}D{.}{.}{4}l}\hline
\multicolumn{1}{c}{Idx} & \multicolumn{1}{c}{SNR} & \multicolumn{1}{c}{$\log_{10}(\GFA)$} & \multicolumn{1}{c}{Segment} &  \multicolumn{1}{c}{Frequency} & \multicolumn{1}{c}{Spindown} &  \multicolumn{1}{c}{$\RAJ$}  & \multicolumn{1}{c}{$\DECJ$} & Description \\
\multicolumn{1}{c}{}	&  \multicolumn{1}{c}{}	&  \multicolumn{1}{c}{}	&  \multicolumn{1}{c}{}	& \multicolumn{1}{c}{Hz}	&  \multicolumn{1}{c}{nHz/s} & \multicolumn{1}{c}{degrees} & \multicolumn{1}{c}{degrees} & \\
\hline \hline
1 & 194.1 & -188.8 & $[0,0]$ & 108.83151 & -3.090 & 323.755 & 37.114 & Induced by loud hardware injection 3\\
3 & 32.1 & -27.3 & $[0,1]$ & 192.55507 & -0.585 & 306.480 & 34.197 & Induced by loud hardware injection 8\\
4 & 31.3 & -36.8 & $[2,5]$ & 69.73917 & -2.885 & 313.580 & 16.478 & Line in L1, Non Gaussian\\
6 & 14.7 & -7.3 & $[2,5]$ & 988.82017 & 0.215 & 307.682 & 33.630 & Line in H1, Non Gaussian\\
7 & 11.4 & -2.9 & $[2,6]$ & 648.74939 & -5.000 & 299.622 & 33.315 & Line in H1\\
8 & 10.8 & -1.2 & $[2,5]$ & 1143.32783 & -0.995 & 300.553 & 32.632 & Strong disturbance in H1\\
9 & 10.2 & 1.4 & $[0,1]$ & 481.96422 & -0.105 & 301.872 & 34.275 & Line in L1, disturbed background in H1 and L1\\
10 & 10.2 & 0.5 & $[2,4]$ & 99.14832 & -1.100 & 318.949 & 27.088 & Disturbed background in L1\\
11 & 9.8 & 0.3 & $[1,4]$ & 897.63729 & -2.215 & 303.190 & 35.780 & Non Gaussian\\
12 & 9.8 & 1.0 & $[2,4]$ & 956.74358 & -4.115 & 302.033 & 34.395 & Disturbed background in H1, Non Gaussian\\
13 & 9.8 & -0.7 & $[1,6]$ & 1138.50993 & 0.090 & 299.389 & 34.748 & Disturbed background in H1+L1, Non Gaussian\\
14 & 9.7 & -0.9 & $[0,6]$ & 1404.89226 & -1.205 & 303.637 & 36.819 & \\
15 & 9.6 & -0.8 & $[0,6]$ & 799.42915 & -0.840 & 300.724 & 31.062 & Line in H1, Non Gaussian\\
16 & 9.5 & 0.1 & $[1,5]$ & 1368.77913 & -3.560 & 304.484 & 30.949 & Lines in H1\\
17 & 9.4 & 2.4 & $[1,2]$ & 1308.96651 & -1.670 & 304.436 & 30.232 & Non Gaussian\\
18 & 9.4 & 0.8 & $[2,5]$ & 1386.45871 & -0.510 & 304.398 & 34.228 & Line in H1 at 1386.5~Hz, Non Gaussian\\
21 & 9.2 & 2.6 & $[5,6]$ & 1170.98217 & -4.395 & 304.353 & 34.829 & Non Gaussian\\
22 & 9.0 & 4.1 & $[2,2]$ & 1191.26642 & -0.455 & 300.720 & 31.494 & \\
23 & 8.9 & 2.1 & $[0,2]$ & 829.72137 & -2.900 & 305.831 & 33.090 & \\
24 & 8.9 & 0.4 & $[0,6]$ & 1321.56703 & -1.820 & 304.707 & 32.001 & Non Gaussian\\
25 & 8.9 & 2.9 & $[4,5]$ & 1058.43325 & -3.600 & 300.356 & 31.068 & \\
26 & 8.9 & 1.6 & $[1,4]$ & 1302.65337 & -2.250 & 299.854 & 34.786 & \\
27 & 8.8 & 0.9 & $[0,5]$ & 1474.94224 & -2.050 & 303.295 & 32.273 & \\
28 & 8.8 & 0.6 & $[0,6]$ & 990.76130 & -2.705 & 299.638 & 33.235 & Disturbed background in H1\\
29 & 8.7 & 1.1 & $[1,6]$ & 1429.67892 & -2.010 & 303.739 & 32.845 & \\
30 & 8.6 & 0.9 & $[0,6]$ & 1325.50969 & -4.325 & 300.291 & 34.313 & Disturbed background in L1, Non Gaussian\\
31 & 8.5 & 3.4 & $[5,6]$ & 1177.15326 & -0.040 & 307.054 & 32.374 & \\
32 & 8.4 & 1.5 & $[1,6]$ & 1330.69434 & -3.285 & 300.625 & 34.037 & Disturbed background in H1, Non Gaussian\\
33 & 8.4 & 1.5 & $[0,5]$ & 1456.26611 & 0.195 & 302.336 & 33.628 & L1 SNR is inconsistent with background level\\
34 & 8.3 & 2.0 & $[2,6]$ & 995.14313 & -1.400 & 302.428 & 31.768 & Disturbed background in L1\\
35 & 8.1 & 1.8 & $[0,6]$ & 1286.17215 & -1.185 & 305.624 & 35.126 & Line in H1, Non Gaussian\\
36 & 8.0 & 2.8 & $[2,5]$ & 1386.02201 & 0.050 & 304.242 & 36.465 & Line in H1 at 1385.9~Hz, Non Gaussian\\
37 & 7.8 & 4.2 & $[1,2]$ & 1359.72387 & -1.745 & 298.903 & 32.885 & Instrumental contamination in L1\\

\hline
\end{tabular}
\caption[Outliers that passed the full detection pipeline for region A]{Outliers that passed the full detection pipeline from region A. Only the highest-SNR outlier is shown for each 0.1~Hz frequency region. Outliers marked with ``line'' had strong narrowband disturbance identified near outlier location. Outliers marked as ``non Gaussian'' were identified as having non Gaussian statistic in their power sums, often due to very steeply sloping spectrum. }
\label{tab:outliersA}
\end{center}
\end{table*}

\begin{table*}[htbp]
\begin{center}
\small
\begin{tabular}{D{.}{.}{2}D{.}{.}{2}D{.}{.}{2}rD{.}{.}{5}rD{.}{.}{4}D{.}{.}{4}l}\hline
\multicolumn{1}{c}{Idx} & \multicolumn{1}{c}{SNR} & \multicolumn{1}{c}{$\log_{10}(\GFA)$} & \multicolumn{1}{c}{Segment} &  \multicolumn{1}{c}{Frequency} & \multicolumn{1}{c}{Spindown} &  \multicolumn{1}{c}{$\RAJ$}  & \multicolumn{1}{c}{$\DECJ$} & Description \\
\multicolumn{1}{c}{}	&  \multicolumn{1}{c}{}	&  \multicolumn{1}{c}{}	&  \multicolumn{1}{c}{}	& \multicolumn{1}{c}{Hz}	&  \multicolumn{1}{c}{nHz/s} & \multicolumn{1}{c}{degrees} & \multicolumn{1}{c}{degrees} & \\
\hline \hline
1 & 41.3 & -55.9 & $[1,4]$ & 243.27113 & -3.675 & 134.486 & -35.443 & Line in H1\\
3 & 19.8 & -20.9 & $[0,6]$ & 69.74870 & -4.130 & 111.634 & -36.471 & Line in L1, Non Gaussian\\
4 & 15.3 & -6.8 & $[1,3]$ & 268.96658 & -5.065 & 135.288 & -46.431 & Line in H1\\
5 & 11.7 & -2.5 & $[2,5]$ & 170.84304 & -2.725 & 124.589 & -48.321 & H1 SNR is larger than coherent sum\\
6 & 11.3 & 0.2 & $[2,3]$ & 108.07698 & -0.115 & 122.585 & -48.207 & Disturbed background in H1\\
7 & 10.9 & -2.0 & $[1,5]$ & 158.39427 & -3.550 & 122.974 & -49.793 & Line in H1\\
8 & 10.8 & 2.5 & $[0,0]$ & 1111.39559 & -0.345 & 131.270 & -44.537 & \\
9 & 10.8 & 2.5 & $[3,3]$ & 956.81519 & -0.905 & 129.372 & -44.282 & Disturbed background in H1\\
10 & 10.6 & 2.7 & $[3,3]$ & 950.80278 & -1.900 & 128.821 & -45.115 & Disturbed background in H1\\
11 & 10.5 & 2.8 & $[0,0]$ & 611.12967 & 0.255 & 130.848 & -49.230 & Non Gaussian\\
12 & 10.0 & -0.6 & $[2,6]$ & 1076.04377 & -3.250 & 133.282 & -47.130 & Line in L1 at 1076~Hz, Non Gaussian\\
13 & 9.9 & 3.3 & $[3,3]$ & 1118.06896 & -2.645 & 128.952 & -47.992 & Disturbed background in L1\\
14 & 9.6 & 0.0 & $[2,6]$ & 1498.30429 & -2.000 & 131.393 & -48.022 & Disturbed background in L1\\
15 & 9.6 & -0.7 & $[0,6]$ & 613.26132 & -3.950 & 125.353 & -42.144 & Non Gaussian\\
16 & 9.3 & 0.4 & $[2,6]$ & 1498.73031 & -0.195 & 125.668 & -42.539 & Non Gaussian\\
17 & 9.3 & 0.0 & $[0,5]$ & 933.33823 & 0.100 & 127.556 & -48.783 & Non Gaussian\\
18 & 9.1 & 1.3 & $[0,3]$ & 1313.24312 & -5.000 & 127.562 & -47.859 & Disturbed background in H1\\
19 & 8.9 & 1.6 & $[0,3]$ & 1458.79267 & -2.425 & 125.394 & -43.661 & \\
20 & 8.9 & 0.8 & $[1,6]$ & 1249.43835 & -1.550 & 128.846 & -46.928 & Disturbed background in H1+L1, Non Gaussian\\
21 & 8.7 & 1.1 & $[1,6]$ & 880.40175 & -2.865 & 130.890 & -47.472 & Disturbed background in H1, Non Gaussian\\
22 & 8.6 & 2.5 & $[2,4]$ & 1254.11705 & -1.295 & 128.862 & -41.615 & Line in L1, Non Gaussian\\
23 & 8.6 & 1.3 & $[1,6]$ & 1333.27906 & -1.650 & 128.265 & -47.879 & Non Gaussian\\
24 & 8.6 & 3.3 & $[1,2]$ & 1497.14217 & -2.210 & 129.202 & -46.366 & Line in H1, Non Gaussian\\
25 & 8.6 & 3.3 & $[2,3]$ & 1333.83095 & -4.445 & 124.636 & -46.522 & Non Gaussian\\
26 & 8.4 & 1.8 & $[0,4]$ & 1336.24255 & -0.005 & 126.374 & -42.633 & Non Gaussian\\
27 & 8.3 & 4.7 & $[1,1]$ & 1370.69201 & -3.375 & 130.199 & -41.568 & \\
28 & 8.3 & 2.1 & $[1,5]$ & 1316.98962 & -1.025 & 130.853 & -47.324 & Non Gaussian\\
29 & 8.2 & 4.9 & $[2,2]$ & 795.42245 & -3.855 & 131.083 & -47.536 & \\
30 & 8.2 & 1.6 & $[0,6]$ & 1458.53648 & -3.800 & 131.684 & -43.218 & Line in H1\\
31 & 8.1 & 2.0 & $[0,5]$ & 1119.11347 & -3.975 & 125.384 & -43.551 & Disturbed background in L1\\
32 & 7.9 & 3.0 & $[2,5]$ & 1331.56844 & -4.210 & 128.647 & -46.135 & Disturbed background in H1\\
33 & 7.6 & 3.8 & $[3,5]$ & 1334.83602 & -4.700 & 129.188 & -41.911 & \\

\hline
\end{tabular}
\caption[Outliers that passed the full detection pipeline for region B]{Outliers that passed the full detection pipeline from region B. Only the highest-SNR outlier is shown for each 0.1~Hz frequency region. Outliers marked with ``line'' had strong narrowband disturbance identified near outlier location. Outliers marked as ``non Gaussian'' were identified as having non Gaussian statistic in their power sums, often due to very steeply sloping spectrum.}
\label{tab:outliersB}
\end{center}
\end{table*}


\begin{table*}[htbp]
\begin{center}
\begin{tabular}{lD{.}{.}{5}rD{.}{.}{2}D{.}{.}{2}}\hline
Name & \multicolumn{1}{c}{Frequency} & \multicolumn{1}{c}{Spindown} & \multicolumn{1}{c}{$\RAJ$} & \multicolumn{1}{c}{$\DECJ$} \\
 & \multicolumn{1}{c}{Hz} & \multicolumn{1}{c}{Hz/s} & \multicolumn{1}{c}{degrees} & \multicolumn{1}{c}{degrees} \\
\hline \hline
ip3   &  108.85716 & $\sci{-1.46}{-17}$   &      178.37  &   -33.44  \\
ip8   & 193.48479 & $\sci{-8.65}{-09}$   &       351.39  &   -33.42   \\
\hline
\end{tabular}
\caption[Parameters of hardware injections]{Parameters of hardware-injected simulated signals detected by PowerFlux (epoch GPS 846885755).}
\label{tab:injections}
\end{center}
\end{table*}


Tables \ref{tab:outliersA} and \ref{tab:outliersB} list outlier index (an identifier used during follow-up), signal-to-noise ratio, decimal logarithm of Gaussian false alarm as computed by formula \ref{GFA}, the contiguous segment of data where the outlier had the highest SNR (see below), frequency, spindown, right ascension and declination, as well as a summary of manual follow-up conclusions.

The segment column describes the persistence of the outlier throughout the analysis.  The data to be analyzed was divided into seven equal-duration segments labeled 0 through 6.  For a continuous signal, the maximum SNR is achieved by integrating all segments: this is indicated by the notation [0,6].  For a transient artifact \cite{transient1}, one can achieve higher SNR by analyzing only those segments when it was on.  This case is indicated by noting the continuous set of segments that gives the largest SNR: e.g. [1,5] if a higher SNR is achieved by dropping the            
first and last segment.  Note, however, that an astrophysical signal such as a long-period binary may also appear more strongly in some segments than others, and thus could have a segment notation other than [0,6]. The same will be true of a strong
signal outside of the search area on the sky, whose Doppler shifts
happen to align with the target area's over some segment of time.
This occurs, for instance, with outliers A1 and A3, which were
generated by a strong simulated signal outside of the search area.


For a low SNR continuous signal it is also possible for the background noise to randomly align in such a way that the SNR over [0,6] segment is slightly lower than on a smaller subset. Our simulations show that $98.5$\% of injections achieve maximum SNR over one of [0,6], [0,5] or [1,6] segments.



Outliers marked as non-Gaussian were found to lie in bands whose statistics deviated from
Gaussian noise, according to the following criterion: the excess kurtosis of 
 $501$ bins around the outlier was smaller than $-1.05$. The probability of Gaussian sample having this excess kurtosis is smaller than $10^{-6}$.


If manual inspection of an
outlier indicated that it overlaps with a strong spectral disturbance
in one of the detectors, this is noted in the tables.  Disturbances
might be either narrow lines, or steep slopes or edges characteristic
of wandering lines or the wings of nearby spectral features.  When
such contamination is manifestly obvious under visual inspection, it
is likely that the outlier was due to that artifact rather than an
astrophysical signal. Outliers with identified contamination are marked with comments in Tables \ref{tab:outliersA} and \ref{tab:outliersB}. 


Two of the outliers were induced by very loud simulated hardware injections. The true parameters of these signals are listed in Table \ref{tab:injections}.

%

\section{Manual outlier followup}

To determine whether or not any of the outliers in Tables \ref{tab:outliersA} and \ref{tab:outliersB} indicated a credible gravitational wave detection, 
each outlier was subjected to manual
inspection, after which several criteria were used to eliminate those
not likely due to the target astrophysical population.  First, we
discarded any candidate with a segment other than [0,6], [0,5], or
[1,6]: as noted, this would eliminate less than 1.5\% of true signals
from our population.  Next, we disregard those signals marked as
"non-Gaussian".  This criterion has a more substantial false dismissal
probability: roughly 20\% of the search band was so marked.
Nonetheless, we would be unable to claim with any confidence that a
candidate from such a band was \emph{not} simply a non-Gaussian
instrumental outlier.  Finally, we disregard outliers in bands with
visually obvious spectral disturbances: this has a similar false
dismissal rate, but has substantial overlap with the non-Gaussian
bands.

\begin{figure}[htbp]
\begin{center}
  \includegraphics[width=3.0in]{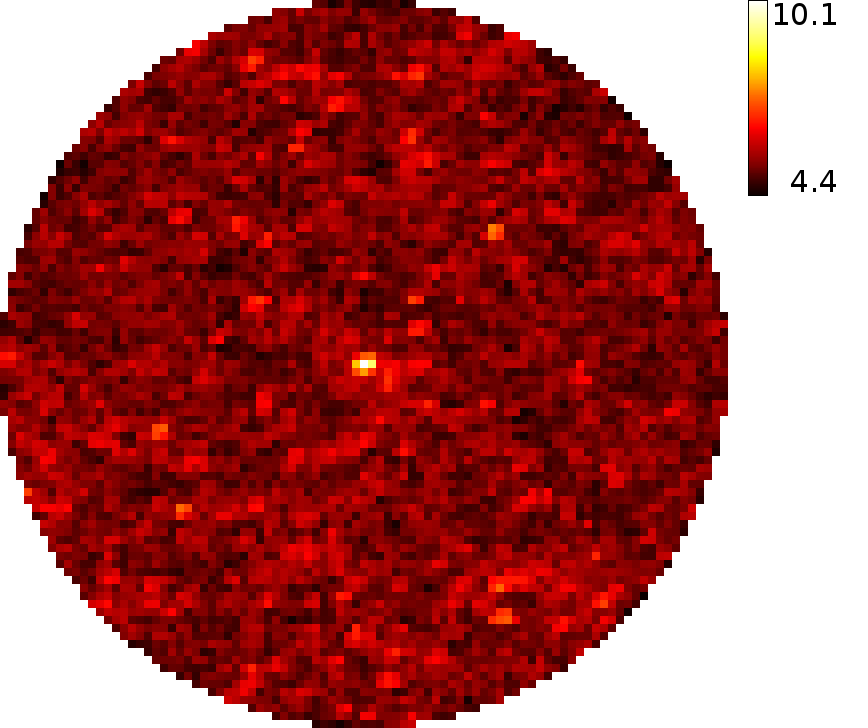}
 \caption[SNR skymap for outlier 14]{SNR skymap for outlier A 14. The  disk ($0.025$~rad radius) is centered on the location of the signal. Each pixel ($0.555$~mrad) on the skymap shows the SNR maximized over a $5\times 5$ template sub-grid and all polarizations. The high frequency of the signal (1404~Hz) allows good localization (color online)}
\label{fig:snr_skymap14_wide}
\end{center}
\end{figure}

This winnowing resulted in three surviving candidates: A14, A27, and
A29.  Of these, A14 is the most interesting (Figure \ref{fig:snr_skymap14_wide}), with a $\log_{10}(\GFA)$ of
$-0.9$.  This suggests that, conservatively, roughly 10\% of searches of
this type would produce an outlier as loud as A14 due to Gaussian
noise alone.  While not enough to make a confident claim of detection,
this was certainly enough to motivate further follow-up.

All three candidates were followed up with NOMAD \cite{miroslav, FullS5EH}, a hierarchical pipeline used in previous
continuous-wave searches \cite{FullS5EH}.  This adaptive pipeline
searched a span of 255 days of S6 data in 5 successive stages of
refinement: with coherent segment lengths of 2.5 days, 5 days, 7.5 days, 10 days, 12.5
days, and 255 days (fully coherent).  The recovered power from each candidate remained
roughly constant at each stage, and consistent with noise, rather than
increasing with coherence length.  This strongly indicates that these
outliers do not follow the presumed signal model over timescales of
several days.

As a consistency check we have also studied the outliers with long coherence codes based on $\mathcal F$-statistic \cite{S2FstatPaper, jks, cutler, miroslav} as well as codes with shorter coherence lengths \cite{radiometer}. The search \cite{PostCasA} with coherence time of 27 days established upper limits at the outlier locations ruling out any significant signals.

\section{Conclusions}

We have performed the first deep search along the Orion spur for continuous gravitational waves 
in the range 50-1500~Hz, achieving a factor of 2 improvement over results from all-sky searches. Exploring a large spindown range, we placed upper limits on both expected and unexpected sources.  At the highest frequencies we are sensitive to neutron stars with an equatorial 
ellipticity as small as $\sci{5}{-7}$ and as far away as $1000$~pc for favorable spin orientations.

A detection
pipeline based on a loosely coherent algorithm was applied to outliers
from our search.  Three outliers (A14, A27, and A29 on Table \ref{tab:outliersA})
were found with continuous presence and no obvious instrumental
contamination.  However, deeper follow-up did not reveal a source
consistent with the original signal model.  This, combined with the
only modest improbability of the loudest outlier occuring in Gaussian
noise, leads us to conclude that statistical fluctuations are the
likely explanation for these outliers.


\section{Acknowledgments}

The authors gratefully acknowledge the support of the United States
National Science Foundation for the construction and operation of the
LIGO Laboratory, the Science and Technology Facilities Council of the
United Kingdom, the Max-Planck-Society, and the State of
Niedersachsen/Germany for support of the construction and operation of
the GEO600 detector, and the Italian Istituto Nazionale di Fisica
Nucleare and the French Centre National de la Recherche Scientifique
for the construction and operation of the Virgo detector. The authors
also gratefully acknowledge the support of the research by these
agencies and by the Australian Research Council, 
the International Science Linkages program of the Commonwealth of Australia,
the Council of Scientific and Industrial Research of India, 
the Istituto Nazionale di Fisica Nucleare of Italy, 
the Spanish Ministerio de Educaci\'on y Ciencia, 
the Conselleria d'Economia Hisenda i Innovaci\'o of the
Govern de les Illes Balears, the Foundation for Fundamental Research
on Matter supported by the Netherlands Organisation for Scientific Research, 
the Polish Ministry of Science and Higher Education, the FOCUS
Programme of Foundation for Polish Science,
the Royal Society, the Scottish Funding Council, the
Scottish Universities Physics Alliance, The National Aeronautics and
Space Administration, the Carnegie Trust, the Leverhulme Trust, the
David and Lucile Packard Foundation, the Research Corporation, and
the Alfred P. Sloan Foundation.

This document has been assigned LIGO Laboratory document number \texttt{LIGO-P1500034-v23}.

\newpage

\end{document}